\documentclass[a4paper,11pt]{article}

\usepackage{vmargin}
\usepackage{amsfonts}
\usepackage{amsmath}
\usepackage{amssymb}  
\usepackage{graphics,graphicx,color}
\usepackage{psfrag}

\newcommand{\beq}{\begin{equation}}
\newcommand{\eeq}{\end{equation}}
\newcommand{\beqa}{\begin{eqnarray}}
\newcommand{\eeqa}{\end{eqnarray}}
\newcommand{\nn}{\nonumber}
\newcommand{\R}{\mathbb{R}}

\newcommand{\la}{\langle}
\newcommand{\ra}{\rangle}

\newcommand{\lalg}[1]{\mathfrak{#1}}  

\newcommand{\SO}{\mathrm{SO}}

\newcommand{\so}{\lalg{so}}

\newcommand{\Aut}{\mathrm{Aut}}
\newcommand{\Ad}{\mathrm{Ad}}
\newcommand{\End}{\mathrm{End}}

\DeclareMathOperator{\tr}{tr}

\newtheorem{lem}{Lemma}

\newtheorem{theo}{Theorem}

\setpapersize{USletter}
\setmarginsrb{39mm}{12mm}{28mm}{8mm}{12pt}{11mm}{0pt}{13mm}		 
    
\begin{document}

\sloppy
\title{\Large\bf Canonical Analysis of Algebraic String Actions}

\author{Winston J. Fairbairn \footnote{winston.fairbairn@nottingham.ac.uk}\  
\\ [1mm]
\itshape{\normalsize{School of Mathematical Sciences, University of Nottingham}} \\
\itshape{\normalsize{University Park, Nottingham, NG7 2RD, UK}} \\
\\ [1mm]
Karim Noui \footnote{Karim.Noui@lmpt.univ-tours.fr} , Francesco Sardelli \footnote{Francesco.Sardelli@lmpt.univ-tours.fr}\
\\ [1mm]
\itshape{\normalsize{Laboratoire de Math\'ematiques et Physique Th\'eorique}} \\
\itshape{\normalsize{Universit\'e Fran\c cois Rabelais, Parc de Grandmont, 37200 Tours, FR}} }
\date{{\small\today}}
\maketitle

\abstract{We investigate the canonical aspects of the algebraic first order formulation of strings introduced two decades ago by Balachandran and collaborators. We slightly enlarge the Lagrangian framework and show the existence of a self-dual formulation and of an Immirzi-type parameter reminiscent of four-dimensional first order gravity. We perform a full Hamiltonian analysis of the self-dual case: we extract the first class constraints and construct the Dirac bracket associated to the second class constraints.  The first class constraints contain the
diffeomorphisms algebra on the world-sheet, and the coordinates are shown to be non-commutative with 
respect to the Dirac bracket. The Hamilton equations in a particular gauge are shown to 
reproduce the wave equation for the  string coordinates.
In the general, non-self-dual case, we also explicit the first class constraints of the system and show that, unlike the self-dual formulation, the theory admits an extra propagating degree of freedom than the two degrees of freedom of conventional string theory. This prevents the general algebraic string from being strictly equivalent to the Nambu-Goto
string.}

\section{Introduction}

Relativistic point particles in Minkowski space can be described in terms of algebraic variables given by coordinates on the Poincar\'e group manifold \cite{Bal1}. The translations describe position while the Lorentz group elements encode the momentum and spin of the particles. The dynamics is governed by an algebraic action constructed using the Maurer-Cartan form on the Poincar\'e Lie algebra. This formulation is valid in arbitrary space-time dimensions, Lorentzian and Euclidean signatures (by using the Poincar\'e group or the Euclidean group), and can also describe extended objects such as strings and branes \cite{Bal2}. When applied to spinless strings in Minkowski space, the framework suggests an appealing classification of string theories as tachyonic, null or of Nambu-Goto type similar to that of relativistic point particles. It can also describe spinning objects \cite{Stern} purely in terms of bosonic variables, without the introduction of supersymmetry.

This alternative first order formulation of strings is interesting for several reasons.
Firstly, the (spinless) action is quadratic in the string coordinates, as the Polyakov action, and is thus better suited for path integral quantisation than the original Nambu-Goto action. Unlike the Polyakov action, the formalism does not depend on any metric structure on the world-sheet and resembles a two-dimensional topological field theory. The action is obtained in purely geometrical terms as the integral of a two-form over a surface, and the target space information is coded algebraically in a choice of a symmetry group, typically given by the inhomogeneous isometry group of a flat metric for the background geometry. This suggests the possibility that the framework could describe strings propagating on other maximally symmetric solutions to Einstein's equations such as deSitter or anti-deSitter spaces.

Secondly, this formulation of strings, or more generally the algebraic formulation of matter, is interesting for its relation to gravity. First, the algebraic formulation of strings offers striking analogies with first order gravity.
It is a diffeomorphism invariant theory with no explicit dependence on a metric structure, it admits an infinite number of local degrees of freedom, and the action is constructed using differential forms taking value in the isometry algebra of a flat metric. As we will see in the core of the paper, this analogy can be precisely formulated. Second, the formalism naturally couples to gravity. For instance, the algebraic description of point particles has led to substantial progress in the quantisation of matter degrees of freedom coupled to 2+1 gravity \cite{Freidel1,effqg,Karimale,Karim}. The underlying reason is that the formalism is based on the Maurer-Cartan form on the Poincar\'e algebra, that is, a flat, pure gauge connection for the Poincar\'e group. The coupling to gravity, regarded as a Poincar\'e gauge theory, then naturally occurs by the standard minimal coupling prescription and matter is understood as a local property of the space-time geometry (see \cite{Stern1} for strings coupled to 2+1 gravity). Even if gravity in higher dimensions is no longer a local theory of the Poincar\'e group, algebraic strings and branes naturally couple to gravity and there is some evidence that this framework is better suited for the coupling of matter to quantum gravity \cite{W}. In particular, strings seem to be a natural source \cite{BFloops,BP,AW,Ale} to couple to four-dimensional quantum gravity using the BF formulation of gravity  as a constrained topological field theory \cite{Pleb}.

Finally, one can view the bosonic string as a ``toy-model'' to test some ideas and techniques introduced in the context of 4d quantum gravity.
Indeed, the bosonic string is certainly one of the most interesting system which reproduces, 
in a simpler framework, some important features of gravity: it is a diffeomorphism invariant theory and admits
an infinite number of local degrees of freedom. Furthermore, the theory is simple enough  to be quantised 
\`a la Fock completely. This is of course very well known since more than three decades 
and was the first step towards string theory. A few years ago, Thiemann \cite{Thiemann} reconsidered the Nambu-Goto string and proposed a quantisation of it using the techniques of loop quantum gravity (LQG) \cite{Ash}. He showed that the LQG techniques, based on background independent quantisation, provides in particular a quantisation
of the bosonic string in any dimensions, i.e., there is no need of critical dimensions for the quantum theory to
be consistent. This result has sparked off some discussions \cite{Policastro} and certainly deserves to be understood deeper. We think that the algebraic formulation of the bosonic string is a better starting point to test the LQG 
techniques than the Nambu-Goto string for it admits a lot of similarities with the Ashtekar-Immirzi-Barbero-Holst formulation \cite{selfdualgravity}, \cite{Holst} of general relativity. It is a first order
formulation and possesses an Immirzi-type parameter. In fact, the main motivation of this article is to open an arena
for a background independent quantisation of the bosonic string and to compare it to the standard Fock quantisation.
Our goal is to pursue the line of research initiated by Thiemann in the context of the algebraic formulation of strings.

For all these reasons, we believe that the classical and quantum aspects of the formalism should be better studied.
Strikingly, the literature on the subject is rather sparse and a part from some work by Stern \cite{Stern}, no study of the canonical aspects of the framework exist to our knowledge. The same remark applies for the quantum aspects of the framework. As a first step towards a canonical quantisation of the theory, this paper is devoted to an analysis of the Hamiltonian formulation.

The results of the paper are as follows.
We start by the Lagrangian aspects and slightly enlarge the original framework. This provides an insight on the theory leading to the discovery of a self-dual formulation and of an Immirzi-type parameter. This enhances the relation with four-dimensional gravity. We then study the canonical analysis of the self-dual action and of the general framework. In the self-dual case, we exhibit all the first class constraints explicitly and compute the Dirac bracket associated to the second class constraints. We then calculate the Dirac algebra of the first class constraints, compute the Hamilton equations of motion and show a non-commutativity of the coordinates with respect to the Dirac bracket. In the general case, we also extract the first class constraints but our analysis shows that there is, in this case, an extra propagating degree of freedom leading to the conclusion that, even if the Lagrangian approach coincides with the Nambu-Goto string, the physical content of the theory is different.

\section{The algebraic formulation of string theory}

\subsection{Preliminaries}

In this paper, we will consider a (non-critical) closed spinless bosonic string propagating on a four-dimensional flat Euclidean manifold $M \cong \R^4$ with metric $\eta=(++++)$. Let $\Sigma \subset M$ be the string world-sheet, i.e., a two-dimensional compact sub-manifold of $M$. The string is described by a pair of fields $\Lambda=(X,g)$ on the string world sheet $\Sigma$; $X$ is the embedding map $X : \Sigma \rightarrow M$, and the field $g$ is a smooth map $g : \Sigma \rightarrow \SO(4)$ valued in the isometry group $\SO(\eta) \cong \SO(4)$ of the flat metric $\eta$.
The extension to Lorentzian signatures is immediate and the generalisation to higher dimensions will be studied elsewhere.

\subsubsection{The algebra $\so(4)$ and related notions}

Let $\pi : \SO(4) \rightarrow \Aut (\R^4)$ denote the vector (fundamental) representation of the isometry group $\SO(4)$ and $(e_I)_{I=0,...,3}$ be a choice of basis of $\R^4$. The induced representation $\pi_*:\so(4) \rightarrow \End (\R^4)$ of the Lie algebra $\so(4)$ is defined by the following evaluations on the elements of the basis $(T_{ab})_{a<b=0,...,3}$ of $\so(4)$:
\beq
\label{vector}
\pi_*(T_{ab})^I_{\;\;J}:= (T_{ab})^I_{\;\;J} = \delta_a^I \delta_{bJ} - \delta_{aJ} \delta_b^I,
\eeq
where $\delta$ is the Kronecker symbol.
We introduce the two $\Ad$-invariant, non-degenerate bilinear forms on $\so(4)$ which are defined by
\beq
\langle T_{ab}, T_{cd} \rangle = \delta_{ac} \delta_{bd} - \delta_{ad} \delta_{bc}, \;\;\;\; (T_{ab}, T_{cd}) =  \epsilon_{abcd},
\eeq
where $\epsilon_{abcd}$ is the four-dimensional Levi-Cevita tensor normalised by $\epsilon_{0123} = 1$. Using the expression for the matrix elements of the generators, it is straightforward to relate the two bilinear forms to the trace in the vector representation $\mathrm{Tr}$ and obtain that $\la , \ra = -\frac{1}{2} \mathrm{Tr}$.

Using the linear Hodge involution map 
$$
\star : \so(4) \rightarrow \so(4) \, ; \;\;\;\; T_{ab} \mapsto (\star T)_{ab} = \frac{1}{2} \epsilon_{ab}^{\;\;cd} T_{cd},
$$
one can show that the two bilinear forms are related as follows
\beq
\forall X, Y \, \in \so(4), \;\;\;\; ( X, Y ) = \langle X, \star Y \rangle = \langle \star X, Y \rangle.
\eeq

Via the eigenspace decomposition of the Hodge involution, the Lie algebra $\so(4)$ splits, as a Lie algebra, into two commuting three-dimensional rotation algebras; the self-dual and anti-self-dual subalgebras
\beq
\label{factorisation}
\so(4) \cong \so(3)_+ \oplus \so(3)_-.
\eeq
By virtue of this factorisation, any element $X$ in $\so(4)$ decomposes as $X=X_+ - X_-$ with $X_{\pm} = 1/2 (\pm X + \star X)$ satisfying $\star X_{\pm} = \pm X_{\pm}$. 
In terms of the generators, the isomorphism is given by
\beq
\label{iso}
T_{\pm a} = \frac{1}{2} \left( \pm T_{0a} + \frac{1}{2} \epsilon_{a}^{\;bc} \,T_{bc} \right),
\eeq
where $(T_{\pm a})_{a=1,2,3}$ generate the $\so(3)_{\pm}$ sub-algebra and $\epsilon_{abc} = \epsilon_{0abc}$ is the three-dimensional Levi-Cevita tensor. 
The self-dual and anti-self-dual generators $(T_{\pm a})_a$ satisfy
\beq
[ T_{\pm a} , T_{\pm b} ] = \epsilon_{ab}^{\;\;\,c }T_{\pm c}, \;\;\;\; \mbox{and} \;\;\;\; [T_{+ a}, T_{- b}] = 0,
\eeq
and the Killing form $\tr_{\pm}$ on $\so(3)_{\pm}$ is defined by
\beq
\tr_{\pm} \, T_{\pm a} T_{\pm b} = \delta_{ab}.
\eeq

In fact, the self-dual/anti-self-dual decomposition is orthogonal in both bilinear forms and the Killing form $\langle , \rangle$ reduces to (one-half times) the Killing form on each one of the two copies:
$$
\la T_{\epsilon a}, T_{\epsilon'b} \ra = \frac{1}{2} \delta_{ab} \delta_{\epsilon \, \epsilon'},
$$ 
with $\epsilon, \epsilon' = \pm$.

\subsubsection{Ingredients for the algebraic string}

The basic building block of the algebraic string action is the rank two tensor $dX \otimes dX$ where the symbol $d$ denotes the exterior derivative on the world sheet $\Sigma$ and the tensor product is on the cotangent bundle $T^* \Sigma$. Indeed, the symmetric part of this tensor can be used to reconstruct the induced metric $h$ on the world sheet :
$$
h = X^* \eta = \eta(dX \otimes dX) = \partial_{\alpha} X^I \partial_{\beta} X^J \eta_{IJ} \, dx^{\alpha} \otimes dx^{\beta},
$$
where the star $*$ is the pull-back map, we have chosen a local basis $(\partial_{\alpha} x)_{\alpha=0,1}$ of the tangent space over the surface $\Sigma$, and the metric is regarded as a map $\eta : \R^4 \otimes \R^4 \rightarrow \R$; $e_I \otimes e_J \mapsto \eta_{IJ}$. In turn, the antisymmetric part of this tensor
$$
d X^{[I} \wedge dX^{J]} = B^{IJ} d^2 x,
$$
with 
\beq
B^{IJ} := \epsilon^{\alpha \beta} \partial_{\alpha} X^{[I} \partial_{\beta} X^{J]},
\eeq
defined in terms of the two-dimensional Levi-Cevita tensor $\epsilon$ normalised by $\epsilon_{01} = 1$, is instrumental in measuring the area $A_{\Sigma}$ of the surface $\Sigma$ in the flat background metric:
\beq
A_{\Sigma} = \int_{\Sigma} d^2 x \, \sqrt{\det \, h}.
\eeq
Indeed, using the vector space isomorphism $\Lambda^2(\R^4) \simeq \so(4)$ between the space of two forms (bivectors) over $\R^4$ and the Lie algebra of $\SO(4)$, one can import the Killing form $\langle , \rangle $ on $\so(4)$ to $\Lambda^2(\R^4)$ and construct the quantity $\la B^2 \ra := \la B, B \ra$ which is related to the square of the area $A_{\Sigma}$ because
\beq
\la B^2 \ra = - \frac{1}{2} \mathrm{Tr} \, B^2 = \frac{1}{2} B^{IJ} B_{IJ} = \det \, h.
\eeq
It seems therefore natural to use the $\so(4)$-valued two-form $dX \wedge dX$ to construct the Nambu-Goto action. However, we would still have to deal with a square root. To obtain a more tractable action, especially regarding quantisation, we introduce an auxiliary variable, analogue of the world sheet metric in the Polyakov formulation, and write a first order action. Let $k$ be a fixed element in $\so(4)$ and consider its image under the inner automorphism given by the (inverse) group adjoint action
\beq
M=\Ad_{g^{-1}}(k) = g^{-1} k g, \;\;\;\; g \in C^{\infty}(\Sigma,\SO(4)).
\eeq
The group variable $g$ will play the role of the auxiliary, first order variable.

\subsection{General action and symmetries}

The algebraic string action depends on the variables $X$ and $g$ and is given by \cite{Bal1}:
\beq
\label{general}
S[X,g] = \int_{\Sigma} \la M, dX \wedge dX \ra = \frac{1}{2} \int_{\Sigma} d^2 x \, M_{IJ} \, B^{IJ}.
\eeq

\subsubsection{The symmetries}

This action enjoys two types of symmetries. It is globally invariant under the Euclidean group $\mathrm{I} \SO(4)=\SO(4) \ltimes \mathbb R^4$, 
and also invariant under two local symmetry groups. 

Under the local left action of the $\SO(4)$ subgroup $G_{k}$ that stabilises the element $k$ in $\so(4)$:
\beq
g \rightarrow h\,g, \;\;\;\; \mbox{for all} \;\;\;\; h \in G_k,
\eeq
the algebraic action \eqref{general} is invariant.
The dimension of $G_k$ depends on  $k$. When $k$ and $\star k$ are linearly independent, i.e., $k$ does not belong to one of the sub-algebras $\so(3)_{\epsilon}$ ($\epsilon =\pm$), 
then $G_k \cong \SO(2)\times \SO(2)$ and its Lie algebra is generated by $(k,\star k)$. On the other hand, if $k$ and $\star k$ are linearly dependent, i.e., $k$ belongs to one of the sub-algebras $\so(3)_\epsilon$, then $G_k \cong \SO(3)_\epsilon\times \SO(2)$, the subgroup $\SO(2)$ being generated by $k$. 

The action is also invariant under the infinite dimensional Lie group $\mathrm{Diff}(\Sigma)$ of diffeomorphisms of the surface $\Sigma$. This invariance is immediate to check because the action is metric-independent and given by the integral of a two-form over a two-dimensional manifold.

\subsubsection{Relation to the Nambu-Goto string}

The objects described by the theory are encoded in the value of the free parameter $k$, or more precisely in the conjugacy class to which $k$ belongs. It can be tuned to describe the Nambu-Goto string. To understand this last point, we perform the variation of the action with respect to the group variable $g$. Using the right variation 
$$
\delta g = g \circ T, 
$$
with $T$ an arbitrary element of $\so(4)$, together with the invariance of the Killing form, we obtain the equations of motion
\beq
\delta S = 0 \;\;\;\; \Leftrightarrow \;\;\;\; [M,B] = 0.
\eeq
Thus, the motion forces $M$ to lie in the centraliser $C(B)$ of $B$ in $\so(4)$. Because of the rank of $\so(4)$, $C(B)$ is a two-dimensional (real) vector space spanned\footnote{Note that $B$ and $\star B$ are always independent because $B$ is a non-degenerate, simple bivector, i.e. $\langle B^2 \rangle \neq 0$ and $(B^2) = 0$.} by $B$ and $\star B$. Therefore, the action is extremal if and only if
\beq
M = \alpha B + \beta \star B, \;\;\;\; \alpha,\beta \in \R.
\eeq
The values of $\alpha$ and $\beta$ are then fixed by the conjugacy class of $k$, or equivalently of $M$. Indeed, the conjugacy class of $M$ in $\so(4)$ is labelled by the two adjoint action invariants constructed from the two Ad-invariant bilinear forms
\beq
\tau^2 = \la M^2 \ra, \;\;\;\; \mbox{and} \;\;\;\; s^2 = ( M^2 ),
\eeq
where we have introduced the notation $\la X^2 \ra = \la X , X \ra$, and $( X^2 ) = ( X , X )$ for all $X$ in $\so(4)$.
Using the fact that $( B^2 ) =0$, i.e., the bivector $B$ is {\em simple}, one can express the two unknowns $\alpha,\beta$ as functions of the invariants $\tau^2$ and $s^2$:
$$
\alpha = \frac{1}{2 \sqrt{\la B^2 \ra}} \left( \sqrt{\tau^2 + s^2} + \sqrt{\tau^2 - s^2} \right), \;\;\; 
\beta = \frac{1}{2 \sqrt{\la B^2 \ra}} \left( \sqrt{\tau^2 + s^2} - \sqrt{\tau^2 - s^2} \right).
$$
Solving the equations of motion for the variable $g$ thus leads to the second order action
\beq
S[X] = C(k) \, \int_{\Sigma} d^2 x \, \sqrt{\la B^2 \ra},
\eeq
where 
\beq
C(k) = \frac{1}{2} \left( \sqrt{\tau^2 + s^2} + \sqrt{\tau^2 - s^2} \right).
\eeq
Thus, we obtain the Nambu-Goto action when the conjugacy class of $k$, which is labelled by $\tau^2$ and $s^2$, is fixed such that
\beq
\label{condition}
C(k) = \frac{1}{2 \pi \alpha'},
\eeq
with $\alpha'$ the Regge slope. The choice made in \cite{Bal1} consists in choosing the class of $k$ such that $\tau^2 = 1 / (2 \pi \alpha')^2$ and $s=0$ which indeed leads to the Nambu-Goto action up to a sign
\beq
S[X] = \pm \frac{1}{2 \pi \alpha'} \, A_{\Sigma}.
\eeq
To conclude this section, let us emphasise that the previous calculations do not show the classical equivalence between the Nambu-Goto string and the algebraic string. They only indicate that the classical solutions of the Nambu-Goto action are contained in the set of solutions
of the algebraic action. The fact that one action $S_1$ reduces to another action $S_2$ solving partially the equations of motion of $S_2$ is not enough to claim the equivalence between the two classical theories. In fact, we are going to show that the algebraic string is generically not equivalent to the Nambu-Goto one.

\subsubsection{Self-dual formulation}

One of the main interests in the study of this formulation of string theory is the similarities with gravitational theories. In the mid-eighties, Ashtekar discovered that general relativity could be entirely described by a self-dual connection \cite{selfdualgravity}. Interestingly, a similar phenomenon happens here. 

The self-dual formulation of the algebraic string is obtained as follows. First, we note that, due to the factorisation (\ref{factorisation})
of $\so(4)$ and the orthogonality with respect to the Killing form between the self-dual and anti-self-dual variables,
the algebraic string action \eqref{general} factorises into two pieces according to \eqref{iso} :
\beq\label{lolo}
S[X,(g_+,g_-)] = \frac{1}{2} \int_{\Sigma} d^2 x \, M_{+ \, IJ} \, B^{IJ}_+ - \frac{1}{2} \int_{\Sigma} d^2 x \, M_{- \, IJ} \, B^{IJ}_-,
\eeq
where the $\SO(4)$ group element $g$ has been decomposed according to the self-dual anti-self-dual decomposition into an element $(g_+,g_-)$ of $\SO(3)_+ \times \SO(3)_-$.

The key point is that, similarly to the case of gravity, only one half of the action, that is, the self-dual {\em or} the anti-self-dual part of the action, is sufficient to describe the full Nambu-Goto dynamics. Indeed, considering, for instance, only the self-dual part of the action (\ref{lolo}) is equivalent to choosing a Lie-algebra element $k$ such that $k_-=0$. If we furthermore chose the conjugacy class of $k$ such that $\tau^2 = s^2 = 1 /2(\pi \alpha')^2$, the algebraic string action reduces to a purely self-dual term
\beq
\label{selfdual}
S[X,g_+] = \frac{1}{2} \int_{\Sigma} d^2 x \, M_{+ \, IJ} \, B^{IJ}_+,
\eeq
but still leads to the Nambu-Goto action in its second order form because equation \eqref{condition} remains satisfied with our choice of $k$. How can that be? This is a simply explained by the fact that the area of the surface $\Sigma$ can be measured using only the self-dual (or anti-self-dual) part of the bivector $B$. The simplicity of $B$ implies that $\la B_+^2 \ra= \la B_-^2 \ra$ which implies in turn that $\det h = 2 \la B_+^2 \ra$. In fact, the self-dual trick can be extended to a more general framework.

\subsubsection{Topological term}

The general framework described above shows that we can also add an extra term to the action without changing the classical properties of the theory by, once again, an appropriate choice of $k$. Consider the generalised action
\beq
\label{immirzi}
S[X,g] = \int_{\Sigma} \la M, dX \wedge dX \ra + \gamma \int_{\Sigma} ( M ,dX \wedge dX ),
\eeq
where $\gamma \in \R$ is a free parameter, analogous to the Immirzi parameter in the Holst formulation of gravity 
\cite{Holst}. In the gravitational context, the analogue of this parameter provides a way to circumvent the problem of the reality
constraints when the parameter $\gamma$ is real and is essential for the loop quantum gravity framework to apply.

It is immediate to see that in the case $\gamma = 1$, we recover the self-dual framework discussed above. For a general value of $\gamma$, the action also describes a first order formulation of the Nambu-Goto string since it can be rewritten as
\beq
S[X,g] = \int_{\Sigma} \la M_{\gamma},  \, dX \wedge dX \ra,
\eeq
with $M_{\gamma} = M + \gamma \star M$, and thus enters the general framework presented above, that is, reduces to the Nambu-Goto action with an appropriate choice of the class of $k$. The correct class is labelled by the values $\tau^2 = (1+\gamma^2)/(2 \pi \alpha')^2$, $s^2 = 2 \gamma/ (2 \pi \alpha')^2$ in the $\gamma<1$ case, and $\tau^2 = (1+\gamma^2)/(2 \gamma \pi \alpha')^2$ and $s^2 = 2 / \gamma (2 \pi \alpha')^2$ in the $\gamma>1$ case.

The above action is the most general algebraic action for the string. It contains the self-dual case for $\gamma=1$ and the original action proposed in \cite{Bal1} when $\gamma = 0$. We will therefore refer to this action as the general action.

\section{Hamiltonian analysis}

As a first step towards the canonical quantisation of the theory introduced above, we perform a full Hamiltonian analysis. As in the gravitational context, the self-dual case is simpler and is strictly equivalent to the Nambu-Goto string. 
Nonetheless, contrary to what happens in gravity, the self-dual framework is not equivalent to the generic case which is shown to contain an extra degree of freedom in the configuration space. 
We therefore start by the canonical analysis of the self-dual action. Then, we tackle the general action which is technically more involved. 

The canonical setting is as follows.
We suppose that the world sheet $\Sigma$ is diffeomorphic to the cylinder and foliate it by a one parameter family of one-dimensional `spatial' manifolds $S_t$, $t \in \R$, each diffeomorphic to the circle, that is, $\Sigma \simeq \R \times S$. Let $x \in [0,2 \pi]$ denote a parametrisation of the circle $S$ and let the configuration variables satisfy $X(t,0) = X(t,2\pi)$ and $g(t,0) = g(t,2\pi)$ for all $t$ in $\R$. 

\subsection{Self-dual case}

The canonical decomposition of the self-dual string action \eqref{selfdual} yields
\beq
\label{canonical}
S = \frac{1}{2} \int_{\R} dt \, \int_{S} dx \, M_{IJ} \, \partial_t X^{[I} \partial_x X^{J]},
\eeq
where we have omitted for simplicity the index $+$ to specify the self-dual components. We will adopt this simplification
in the whole section as there is no possible confusion.
From this canonical action, we can read out the momenta conjugate to the configuration variables $(X,g)$ and study the constraints of the system. 

\subsubsection{Symplectic structure}
We start by introducing the momenta $\pi_I$, $I=0,...,3$ conjugate to the variables $X^I$. The corresponding symplectic structure is read out of the Poisson brackets
\beq
\{\pi(x)_I, X(y)^J \} = \delta_I^J \, \delta(x,y).
\eeq

The second configuration variable $g$ is valued in the self-dual $\SO(3)_+$ subgroup of $\SO(4)$. 
The construction of the corresponding symplectic structure enters the general geometrical framework of symplectic structures on co-tangent bundles of Lie groups (see e.g. \cite{Symp}) that we are going to briefly recall here. 

Let $(r^i)_{i=1,...,\mathrm{dim} \, \mathfrak{g}}$ be a set of local coordinates on the Lie group $G$ with Lie algebra $\mathfrak{g}$ in the neighbourhood of a point $g$ in $G$. The coordinates of an element $(g,p_g)$ of the co-tangent space $T_g^*(G)$ are written $(r^i, p_i)$, where $p_g = p_i dr^i$. In this coordinate-dependent language, the canonical symplectic two-form $\omega = \delta p _i \wedge \delta r^i$, with $\delta$ the differential on $T^*(G)$, induces the following Poisson structure
\beq
\{r^i, r^j \} = \{p_i, p_j \} = 0, \;\;\;\; \mbox{and} \;\;\;\; \{r^i, p_j \} = \delta^i_j.
\eeq
As a first step towards a coordinate-free formulation, we introduce the left-action of $G$ on itself
$$
L : G \times G \rightarrow G, \;\;\;\; (h,g) \mapsto L(h,g) = h \, g,
$$
and consider the partial mapping $\chi_g : G \rightarrow G$ obtained from the left action at $g \in G$ fixed, i.e., $\chi_g = L(.\,,g)$. This map induces two linear maps: a push-forward $\chi_{g *}$ from the tangent space $T_h(G)$ to the tangent space $T_{hg}(G)$, and a pull-back $\chi_{g}^*$ from the co-tangent space $T_{hg}^*(G)$ to the co-tangent space $T_{h}^*(G)$. Using the co-tangent map, we can define the (global) left-trivialisation of the co-tangent bundle
\beq
\lambda : T^*(G) \rightarrow G \times \mathfrak{g}^*, \;\;\;\; (g,p_g) \mapsto (g,P = - (\chi_{g}^*)_{h=e}(p_g) ),
\eeq
where the pull-back map $(\chi_{g}^*)_{h=e} : T_g^*(G) \rightarrow T_e^*(G)$ has been evaluated at the identity element $e$ of $G$. The minus sign is a simple matter of convention.
If $(T_a)_{a=1,...,\mathrm{dim} \, \mathfrak{g}}$ denotes a choice of basis of $\mathfrak{g}$ and $(T^a)_{a}$ is the associated dual basis of $\mathfrak{g}^*$, i.e., $T^a(T_b) = \delta^a_b$, we obtain, expressing $P = P_a T^a$ in the dual basis, the useful expression
\beq
\label{P}
P_a = - p_g(T^L_a) = - p_i \, \chi^i_{\;\,a},
\eeq
where $\chi^i_{\;a}$ is the matrix of the linear map $(\chi_{g *})_{h=e}$ relative to the bases $(T_a)_a$ and $(\partial_{i})_i$. This map associates a vector field $\lambda^L(g)$ in $T_gG$ to each Lie algebra element $\lambda$ in $\mathfrak{g}$:
$$
(\chi_{g *})_{h=e}: \mathfrak{g} \rightarrow T_gG, \;\;\;\; \lambda \mapsto \lambda^L(g) = \lambda^a \chi^i_{\;\,a} \partial_{i}.
$$
The vector field $\lambda^L(g)$ is called the left fundamental vector field associated to $\lambda$. An important fundamental vector field is $T^L_a(g)$ which is associated to the basis element $T_a$ of $\mathfrak{g}$. It is called the left frame of $G$ and maps co-vectors at $g$ to elements of $\mathfrak{g}^*$ (i.e. $T^L_a(\omega_i dr^i) = \chi^i_{\;a} \omega_i= \omega_a$). 

From equation \eqref{P} and the expression of the canonical symplectic form on $T^*G$, we can read out the canonical symplectic structure on $T^*G$ in the left-trivialisation (see e.g. \cite{WZW}):
\beq
\omega = - \frac{1}{2} f_{\;\, bc}^{a} \, \chi^b_{\;\,i} \, \chi^c_{\;\,j} P_a \, \delta r^i \wedge \delta r^j + \chi^a_{\;\,i} \, \delta r^i \wedge \delta P_a,
\eeq
where we have used the Maurer-Cartan structure equation of the left co-frame
$$
\delta \chi^a_{\;\,i} = - \frac{1}{2} f_{\;\, bc}^{a} \, \chi^b_{\;\,i} \, \chi^c_{\;\,j} \delta r^j,
$$
with the co-frame related to the frame by $\chi^a_{\;\,i} \chi^i_{\;\,b} = \delta^a_b$ and $f_{ab}^{\;\,c}$ being the structure constants of $\mathfrak{g}$. Using the following matrix identity
\beq
\left[ \begin{array}{cc} 
A & B \\ - B & 0
       \end{array}
\right]^{-1} = \left[ \begin{array}{cc} 
0 & -B^{-1} \\ B^{-1} & B^{-1} A B^{-1}
       \end{array}
\right],
\eeq
it is immediate to invert the above two-form regarded as an anti-symmetric matrix. This procedure yields the corresponding Poisson bivector from which the following Poisson structure follows
\beq
\{r^i, r^j \} = 0, \;\;\;\; \{r^i, P_a \} = - \chi^i_{\;\,a}, \;\;\;\; \{P_a, P_b \} = f_{ab}^{\;\,c} \, P_c.
\eeq 
The last bracket shows that the Poisson structure on $T^*G$ reduces to the standard Kirillov-Kostant Poisson structure on $\mathfrak{g}^*$.

The above formulae are not yet satisfactory since they depend explicitly on coordinates. We introduce the notation $P_{\lambda} = P(\lambda) = \lambda^a P_a$, for all $\lambda$ in $\mathfrak{g}$. Our goal is now to work explicitly with group elements $g$ and not their coordinates. The bracket 
\beq
\{ r^i, P_{\lambda} \} = - \lambda^a \, \chi^i_{\;\, a} \,  = - (\lambda^L)^i  \nn \\
\eeq
is the local, coordinate expression for
\beq
\{ g , P_{\lambda} \} = - \lambda^L(g) =  \lambda \, g,
\eeq
where we have used the relation between the left fundamental vector field $\lambda^L$ on $G$ and the corresponding $\lambda$ in $\mathfrak{g}$:
$$
\lambda^L(g) = \frac{d}{dt} \left( e^{-t \lambda} \, g \right)_{t=0} = -\lambda \, g.
$$ 
As a result, we obtain the following coordinate-free Poisson structure
\beq\label{su2symplectic}
\{g_1, g_2 \} = 0, \;\;\;\; \{P_{\lambda}, g \} = -\lambda \, g, \;\;\;\; \{P_{\lambda_1}, P_{\lambda_2} \} = P_{[\lambda_1,\lambda_2]}.
\eeq
The momentum $P_\lambda$ is in fact the generator of the left derivative on the space of smooth functions on $G$.
One could have also parametrised the symplectic structure using the generator of the right derivative $Q_\lambda$ 
instead. It is related to the previous one by the relation $Q_\lambda:= Ad_g(P_\lambda)=P_{g\lambda g^{-1}}$. For our
purposes, and considering our conventions, it is technically more interesting to use the left derivatives.

Reducing the above framework to the $\SO(3)_+$ subgroup of $\SO(4)$ provides the framework necessary for this paper. We will consider the canonical pair $(g,P_a)_{a=1,2,3}$ with only non-vanishing Poisson brackets given by
\beq
\{P_a(x), g(y)^I_{\;J} \} = -(T_a \, g(x))^I_{\;J} \, \delta(x,y), \;\;\;\; \{P_{a}(x), P_{b}(y) \} = \epsilon_{ab}^{\;\;\,c} \; P_c\, \delta(x,y).
\eeq

\subsubsection{The set of constraints}

The canonical action \eqref{canonical} defines a constrained system since the conjugate momenta are fixed by the following set of primary constraints
\beqa
\label{constraints}
C_I &:=& \pi_I - \frac{1}{2} M_{IJ} \, d X^J \approx 0 \nn \\
\phi_a &:=& P_{a} \approx 0,
\eeqa
where the symbol $d$ denotes the partial derivative on the spatial circle $S$, $d \equiv \partial_x$. The second set of constraints appears because there are no time derivatives of the group variable $g$ in the action. 
These primary constraints satisfy the following Poisson brackets:
\beqa\label{systemofsecondclass}
\{ C_I(x), C_J(y) \} &=& - \frac{1}{2} d M_{IJ}(x) \delta(x,y) \nn \\
\{ C_I(x), \phi_a(y) \} &=& \frac{1}{2} (g^{-1}[T_{a},k]g)_{IJ} d X(x)^J \delta(x,y) \nn \\
\{ \phi_a(x), \phi_b(y) \} &=& \epsilon_{ab}^{\;\;\,c} \; \phi_c(x) \, \delta(x,y).
\eeqa
Here we have simply used the Poisson structure displayed above, a smearing by test functions to obtains the first equality, and the identity $\{P_{a,} M \} = g^{-1}[T_{a},k]g$ with the commutator on the space of four by four matrices. Thus, even if the constraints $\phi_a$ weakly commute between themselves, they do no with the $C_I$ constraints and are accordingly not first class. The same is true for the $C_I$'s. Before extracting the first class constraints, we need to make sure that the Dirac algorithm is closed, that is, that there are no further secondary constraints.

For that purpose we introduce the total Hamiltonian of the theory.
First, we note that the canonical Hamiltonian $H_c$ vanishes weakly, which is expected since we are working with a diffeomorphism invariant theory. We introduce arbitrary Lagrange multipliers $\mu \in C^{\infty}(S,\R^4)$ and $\nu \in C^{\infty}(S,\so(3)_+)$ to write down the total Hamiltonian 
\beq
H_T = \int_{S} dx \, \left( \mu^I C_I + \nu^a \phi_a \right).
\eeq
The total Hamiltonian dictates the temporal evolution of the dynamical variables. It is immediate to see that the conservation in time of the primary constraints
$$
\dot{C}_I = \{H_T, C_I \} \approx 0, \;\;\;\; \mbox{and} \;\;\;\; \dot{\phi}_a = \{H_T, \phi_a \} \approx 0,
$$
does not introduce secondary constraints. Rather, it imposes some constraints on the Lagrange multipliers
\beqa\label{lagrangemultipliers}
d M_{IJ} \mu^J  - \nu^a (g^{-1}[T_{a} , k]g)_{IJ} \, d X^J &\approx& 0 \nn \\
\mu^I (g^{-1}[T_{a} , k]g)_{IJ} \, d X^J &\approx& 0,
\eeqa
where the last equality only holds weakly. Thus the only constraints of the system are the primary constraints given in \eqref{constraints}.

\subsubsection{Constraints analysis and Dirac bracket}

To complete the canonical analysis, we need to separate the first class and second class constraints. 
There are many ways to do so, one of these is based on the resolution of the previous system (\ref{lagrangemultipliers})
with the Lagrange multipliers as unknown. We do not use this technique here, we prefer to guess the number of
first class constraints and try to extract them out of the whole constraints from physical arguments.

Since the action \eqref{canonical} is invariant under diffeomorphisms of the circle, time reparametrisation and the $\SO(2)$ subgroup of $\SO(3)$ stabilising $k$, we expect to find $3$ first class constraints out of the $7$ constraints \eqref{constraints}. This leaves $4$ second class constraints and yields $2 \times 7 - 2 \times 3 - 4 = 4$ physical degrees of freedom for the string, as expected. 

From (\ref{systemofsecondclass}), we see that the functions $(C_I)_I$ form a set of second class constraints. 
However, it is a priori more difficult to extract the strong first class constraints which strongly commute with any
second class constraints. In order to circumvent this difficulty and also
to strongly set the second class constraints to zero, we construct the Dirac bracket with respect to the set $(C_I)_I$. 
To this aim, we need to invert the Dirac matrix
\beq
D_{IJ}(x,y) = \{ C_I(x), C_J(y) \},
\eeq
where $D_{IJ}(x,y) = D_{IJ}(x) \delta(x,y)$, with $D_{IJ}(x)= - \frac{1}{2} d M_{IJ}(x)$. The four-by-four matrix $D_{IJ}(x) \equiv D_{IJ}$ is clearly invertible since it lies purely in the self-dual component of (the vector representation of) $\so(4)$. In fact, we can prove the following lemma

\begin{lem}
The Dirac matrix $D$ is invertible and its inverse $D^{-1}$ is proportional to $D$:
\beq
D^{-1} = c \, [g^{-1}dg , M], 
\eeq
where the coefficient $c$ is given by $c = 2 (2 \pi \alpha')^2 \left(\tr (dg g^{-1})^2 - 2 (\tr dg g^{-1} \, T_{+3})^2 \right)^{-1}$.
\end{lem}

{\em Proof.} We start by showing that the square of the Dirac matrix $D$ is proportional to the identity.
Using the expression \eqref{vector} for the matrix elements of the $\so(4)$ generators $T_{ab}$ in the vector representation and the explicit isomorphism \eqref{iso} between $\so(4)$ and $\so(3)_+ \oplus \so(3)_-$, it is straight forward to work out the matrix elements of the self-dual and anti-self-dual generators
$$
\pi_*(T_{\pm a})^{IJ} = \frac{1}{2} \left( \pm (\delta^{0I} \delta_a^{J} - \delta^{0J} \delta_a^{I}) + \epsilon_{0a}^{\;\;\;IJ} \right).
$$
From this expression
one can show that the image of the self-dual (and anti-self dual) generators in the vector representation satisfy the following relation
\beq
\label{identity}
\pi_*(T_{\pm a}) \circ \pi_*(T_{\pm b}) = -\frac{1}{2} \epsilon_{ab}^{\;\;\;c} \, \pi_*(T_{\pm c}) - \frac{1}{4} \delta_{ab} 1\!\!1.
\eeq
Note that the left hand side is not necessarily an antisymmetric matrix since the term is quadratic and thus in $U(\so(4))$ not in the Lie algebra $\so(4)$. From this, it is immediate to see that the square of the Dirac matrix $D$ is proportional to the identity and therefore
$$\label{squarerelation}
D^2 = D_a D_b \, \pi_*(T_{+ a}) \circ \pi_*(T_{+ b}) = -\frac{1}{4} \tr D^2 \, 1\!\!1,
$$
where the symbol $\tr \equiv \tr_+$ refers to the Killing form on $\so(3)_+$ (which should not be confused with the trace $\mathrm{Tr}$ in the vector representation of $\so(4)$). The above relation 
implies that the inverse of the Dirac matrix is given by
\beq
D^{-1} = a \, D, \;\;\;\; \mbox{with} \;\;\;\; a = -4 / \tr D^2 .
\eeq
To finish the proof, we need to evaluate the matrix $D$ and its norm. Since $D = -\frac{1}{2} d M$, the following equality holds
\beq
\label{com}
D = \frac{1}{2} [g^{-1}dg, M] = \frac{1}{2}g^{-1}[dgg^{-1},k]g\,.
\eeq
From the above expression, we can compute the norm of $D$ as follows
\beqa
\tr D^2 &=& \frac{1}{4} \tr [ d g g^{-1}, k ] \, [d g g^{-1}, k ] \nn \\
          &=& \frac{1}{4} \left( \tr (d g g^{-1})^2 \, \tr k^2 - (\tr (d g g^{-1}) \, k)^2 \right),
\eeqa
where we have explicitly used the relation $\tr [a,b]^2 = \tr a^2 \tr b^2 - (\tr ab)^2$.
We finally recall the fact that $k$ labels the appropriate conjugacy class for the formalism to reproduce the Nambu-Goto framework. We pick $k = (1/\pi \alpha') T_{+3}$ which implies $\tr k^2 = 1/2 (\pi \alpha')^2$ as required. This closes the proof of Lemma $1$ $\square$.

Now, we have constructed all the ingredients to have an explicit form of the Dirac bracket between any pair of functions
$F$ and $G$ on the phase space:
\begin{eqnarray}\label{defofdirac}
\{F,G\}_D \; = \; \{F,G\} \, - \, \{F,C_I\}(D^{-1})^{IJ}\{C_J,G\} \,.
\end{eqnarray}
In particular, the Dirac brackets between the configurational variables are summarised in the following lemma.
\begin{lem}
The Dirac bracket between the configuration variables are all vanishing except for the bracket between the coordinate variables
\beqa
\{g^1, g^2\}_D &=& 0 \nn \\
\{g, X^I\}_D &=& 0 \nn \\
\{X^I,X^J \}_D &=& (D^{-1})^{IJ}.
\eeqa
Therefore, the coordinates of the world-sheet in the target space are non-commuting quantities.
\end{lem}

{\em Proof.} A simple implementation of the definition of the Dirac bracket (\ref{defofdirac}) leads 
immediately to the result above $\square$.

Let us emphasise that, in the algebraic formulation of the string, the phase space is parametrised
by non-commutative coordinates. The non-commutativity obtained is similar to the non-commutativity \`a la Moyal for
the matrix coefficients $D^{-1}$ are central in the Dirac algebra and then could be naively considered as constant.
However, the physical phase space of the algebraic string does not reduce to the Moyal one. The discrepancies come
when we consider the symmetries of the theory generated by the first class constraints we are going to discuss soon.
Indeed, the physical phase space is obtained after the implementation of the first class constraints which have a 
non-trivial action on the string coordinates $X^I$ and also on the matrix elements of $D^{-1}$. This prevents $D^{-1}$
to be considered as a pure constant.

\subsubsection{First class constraints and diffeomorphisms invariance}

Replacing the former Poisson structure by the new Dirac structure, we can solve the constraints $(C_I)_I$ strongly. We are now left with the three constraints $(\phi_a)_{a}$. To proceed, we need to compute their Dirac brackets 
\beqa
\label{dirac}
\{ \phi_a(x) , \phi_b(y) \}_D &=& [\phi_a, \phi_b](x) \delta(x,y) \nn \\
&&- \int_{S^2} dz dt \{ \phi_a(x) , C_I(z) \} (D(z,t)^{-1})^{IJ} \{ C_J(t) , \phi_b(y) \} \nn \\
&=& [\phi_a , \phi_b](x) \delta(x,y) - \frac{c}{4} \, \Delta_{ab \, IJ}(x,y) d X^I(x) d X^J(y),
\eeqa
where the `central extension' is defined via the quantity
\beq
\Delta_{ab \, IJ}(x,y) = \left( g^{-1}[T_{a},k] \, [dg g^{-1} , k] \,  [T_{b},k]g \right)_{IJ}(x) \delta(x,y). 
\eeq

\begin{theo} {\bf (First class constraints.)}\label{firstclasslemma}
The constraints $\phi_a$ are first class with respect to the Dirac bracket 
and therefore generate the local symmetries of theory.
\end{theo}

{\em Proof.} To prove this lemma, we need to show that each remaining constraint $\phi_a$ weakly commutes in the Dirac structure with all the others. 
From equation \eqref{dirac}, we read that this is equivalent to checking that the symmetric part of the matrix valued function $\Delta_{ab \, IJ}$ vanishes. It is sufficient to show that $\Delta_{ab \, IJ}$ weakly vanishes but we will
see that it vanishes strongly.

The proof of this fact relies on the fact that the symmetric part of a matrix of the schematic form $u \, v \, w$, with $u$, $v$ and $w$ being four-by-four representation matrices of $\so(3)_+$ is proportional to the identity. In fact, using \eqref{identity}, one can show that 
$$
(u \, v \, w)_{(IJ)} = \frac{1}{8} \left( (u \times v ) \cdot w  \right) \, \delta_{IJ}.
$$
Here, the symbol $(IJ)$ denotes normalised symmetrisation. We are using the isomorphism $\so(3) \cong \R^3$ to regard the elements $u,v$ and $w$ as three-vectors; the $\so(3)$ Lie algebra structure is mapped onto the $\R^3$ cross product and the Killing form becomes the Euclidean inner product `$\cdot$'. Introducing the notation $x=k$, $y=dg g^{-1}$ and $e_a=T_{+a}$, we can use the above formula to determine the symmetric part of the matrix $\Delta$ via the substitutions $u = e_a \times x$, 
$v=y\times x$ and $w=e_b \times x$. Using the fact that $(e_a \times x) \cdot x = 0$, it is then immediate to see that 
$(u \times v ) \cdot w =0$. Therefore, since $M_{(IJ)} = (g^{-1} M g)_{(IJ)}$ for all four-by-four matrix $M$,
the $\phi_a$ are first class constraints and satisfy the algebra:
\beqa\label{algebraoffirstclass}
\{ \phi_a(x) , \phi_b(y) \}_D &=& \{ \phi_a(x) , \phi_b(y) \} \, = \, \epsilon_{ab}^{\;\;\,c} \; \phi_c(x) \, \delta(x,y).
\eeqa
This closes the proof of Theorem \ref{firstclasslemma} $\;\;\; \square.$
\medskip

We have shown that the first class constraint algebra is isomorphic to a local $\so(3)$ Lie algebra and therefore the 
algebraic string admits an $\so(3)$ local symmetry. We would have expected to obtain the  Witt algebra instead, i.e.,
the algebra of the diffeomorphisms on the world sheet. In fact, the Witt algebra is contained, in a subtle way, in the local $\so(3)$ algebra. Let $\phi(v) = v^a \phi_a = \tr v\phi$ be the evaluation of the linear form $\phi$ on a $\so(3)$-valued function $v$ on the string. The diffeomorphisms correspond to special values of $v$, i.e., particular linear combinations of the constraints. There are three natural choices
for vectors $v$ in the theory: the first one is the constant vector $k$ which is central in the definition of the
algebraic string. The second one is $\mu=dgg^{-1}$ which is the right-invariant Maurer-Cartan form on $\so(3)_+$ (equivalently one could have chosen the left-invariant one $g^{-1}\mu g=g^{-1}dg$), which can be identified with a flat connection on the string. The third one is obtained by taking the cross product between the two previous vectors, $[k,dgg^{-1}]$.
Each of these vectors are of great interest concerning the question of the symmetries and the relation to diffeomorphisms. To make these aspects more concrete, let us define and study the following constraints:
\beqa\label{localbasis}
H \; = \; \phi(k) \;\;\;\;, \;\;\;\;
H_1 \; = \; \phi(\mu) \;\;\;\;, \;\;\;\;
H_0 \; = \; \phi([k,\mu])\;.
\eeqa
The functions $H$, $H_0$ and $H_1$ form a local basis of the set of first class constraints.
They form a Poisson algebra inherited from (\ref{algebraoffirstclass}). As we are going to see,
the algebra structure of these constraints make clear the relation between the local $\so(3)$
symmetry and the diffeomorphisms.

\begin{theo}{\bf (Symmetry algebra.)}
\label{symmetry}
Let us introduce the smeared constraints
$$
H(\alpha) \; = \; \int_S dx \,\alpha(x) \,H(x), \;\;\;\; H_1(u) = \int_S dx \, u(x) H_1(x), \;\;\;\;
 H_0(v)=\int_S dx \,v(x) H_0(x)
$$
with $\alpha, u$ and $v$ in $C^{\infty}(S,\R)$ arbitrary functions independent of the dynamical variables 
of the theory. They satisfy the following Dirac algebra:
\beqa
&&\{H(\alpha),H(\alpha')\}_D \, = \, 0 \, , \,\, 
\{H(\alpha),H_1(u)\}_D \, = \, -H(u d\alpha) \, , \,\, 
\{H(\alpha),H_0(v)\}_D \, = \, 0 \nn \\ 
&&\{H_1(u),H_1(u')\}_D \, = \, H_1(udu'-u'du)\label{H1H1} \, , \,\,
\{H_1(u),H_0(v)\}_D \, = \, H_0(udv - vdu)\label{H1H0} \nn \\
&&\{H_0(v),H_0(v')\}_D \, = \,-\frac{1}{(2\pi\alpha')^2} H_1(vdv'-v'dv) + \tilde{H}(vdv'-v'dv)\label{H0H0}, \nn 
\eeqa 
where $\tilde{H}(x)=H(x) \tr(k dg g^{-1})$ and $\tilde{H}(v)=\int_S dx\,v(x)\tilde{H}(x)$ the associated smeared function.
\end{theo}

{\em Proof.} We now perform the explicit calculations. We will use the fact that
 $\{\phi(\alpha),\phi(\beta)\}_D =\{\phi(\alpha),\phi(\beta)\} $ for any $\so(3)_+$-valued field $\alpha$
and $\beta$ which depends on $g$ explicitly. 
Furthermore, for purposes of clarity and simplicity, it will be useful to introduce the so-called universal notation:
\beqa
&& a_1=a\otimes 1 ,\,\,\,\,\,\text{and} \,\,\,\,\, a_2=1\otimes a ,\,\,\, \nn \\
&& \text{tr}_1(a_1b_2)=\text{tr}(a) b ,\,\,\, \text{tr}_2(a_1b_2)=\text{tr}(b)a 
,\,\,\,\text{tr}_{12}(a_1b_2)=\text{tr}(a)\text{tr}(b) \nn
\eeqa
for $a$ or $b$ in the enveloping algebra $U(\so(3))$. In that framework, the $\so(3)$ Casimir tensor is denoted
$t_{12}=\delta^{ab}T_{a}\otimes T_{b}$ and satisfies the fundamental defining relation:
$$
\text{tr}_1(t_{12}a_1b_2)\, = \, \text{tr}_2(t_{12}a_1b_2) \, = \, ab \,.
$$ 
Finally, the $\so(3)$ symplectic structure (\ref{su2symplectic}) is translated in that language as follows:
$$
\{g_1,g_2\}=0 ,\,\,\,\{P_1,g_2\}=-t_{12} g_2,\,\,\, \{P_1,P_2\}=[t_{12},P_1]
$$
where $P$ is identified with a $\so(3)$ element and is related to $P_\lambda$ by $P_\lambda=\text{tr}(\lambda P)$
for any $\lambda \in \so(3)$. One can also deduce the formula $\{P_1,g_2^{-1}\}=g_2^{-1}t_{12}$ for the action 
of the left derivative on the inverse group element.

To prove the theorem, we will make use of the following intermediate result:
\beqa\label{intermediateresult}
\{P_1(x),\mu_2(y)\} & = & \{P_1(x),\partial_yg_2(y)\}g_2^{-1}(y) + \partial_yg_2(y) \{P_1(x),g_2^{-1}(y)\}\nn \\
& = & -t_{12} g_2(x)g_2^{-1}(y) (\partial_y \delta(x,y)) \, + \, \mu_2(x) t_{12} \delta(x,y) 
\eeqa
where $\delta(x,y)$ is the delta distribution on the string.

Now, we are ready to perform the calculation of each Poisson bracket.

\vspace{3mm}
$\bullet$ \underline{$\{ H(\alpha) , H(\beta) \}$:}
\vspace{1mm}
\beqa
\{ H(\alpha) , H(\alpha') \} &=& \text{tr}_{12} \int dx dy \, \alpha(x) \alpha'(y) \, k_1k_2\{P_1(x) , P_2(y) \} \nn \\
                           &=& \int dx \, \alpha(x) \alpha'(x) \, \text{tr}([k,k]P(x)) =0 \,.\nn
\eeqa

\vspace{3mm}
$\bullet$ \underline{$\{ H(\alpha) , H_1(u) \}$:}
\vspace{1mm}

\begin{eqnarray}
\{ H(\alpha) , H_1(u) \} &=& \text{tr}_{12}\int dx dy \, \alpha(x) u(y) \, k_1 \{P_1(x),P_2(y) \mu_2(y) \} \nn \\
 & = & \text{tr}_{12}\int dx dy \, \alpha(x) u(y) \, k_1 \left(\{P_1(x),P_2(y)\} \mu_2(y) \right. \nn \\
&& \left. +P_2(y) \{P_1(x),\mu_2(y)\} \right). \nn 
\end{eqnarray}
Here we use the result (\ref{intermediateresult}), the fundamental relation of the Casimir tensor and
we see that the Poisson bracket is a sum of three terms:
\beqa
\{ H(\alpha) , H_1(u) \} &=& \text{tr}_{12}\int dx dy \, \alpha(x) u(y) \,k_1
\left([t_{12},P_1(x)]\mu_2(x) \right. \nn \\ && \left. + \, P_2(x) \mu_2(x) t_{12} \right) \delta(x,y) \nn \\
&&-\text{tr}_{12}\int dx dy \, \alpha(x) u(y) \,k_1 g_2(x) g_2^{-1}(y) t_{12} \, \partial_y \delta(x,y) \nn \\
 &= & \int dx dy \, \alpha(x) u(y) \, 
 \left(\text{tr}(k[\mu(x),P(x)]) \, + \, \text{tr}(kP(x)\mu(x)) \right) \delta(x,y) \nn \\
& & \, - \, \int dx dy \, \alpha(x) u(y) \, \text{tr}(P(y)kg(x)g^{-1}(y))\, \partial_y \delta(x,y). \nn
\eeqa
At this level, we perform an integration by part with respect to the $y$ variable in the third term, denoted hereafter
$I$, which becomes after the integration of the delta distribution:
\beqa
I & = & \int dx \, \alpha(x) \,\text{tr}(kg(x) \partial_x(u(x) g^{-1}(x) P(x))) \nn \\
  & = & -\int dx \, u(x) \, \text{tr}(k \partial_x(\alpha(x)g(x)) g^{-1}(x) P(x)) \nn \\
  & = & -\int dx \, \left( \alpha(x) u(x) \text{tr}(k\mu(x) P(x)) \, - \, u(x) \partial_x \alpha(x) \text{tr}(kP(x)) \right).\nn
\eeqa
Combining this result with the two other terms in the expression of the Poisson bracket of interest,
we finally obtain:
\beqa
\{ H(\alpha) , H_1(u) \} \, = \, - \int dx \, \partial\alpha(x) u(x) \, H(x) \, = \, -H(u d\alpha),
\eeqa
as proposed in the theorem.

\vspace{3mm}
$\bullet$ \underline{$\{ H(\alpha) , H_0(v) \}$:}
\vspace{1mm}

\beqa
\{ H(\alpha) , H_0(v) \} &=& \text{tr}_{12}\int dx dy \, \alpha(x) v(y) \, k_1 \{P_1(x), P_2(y) [k_2,\mu_2(y)]\} \nn \\
 & = & \text{tr}_{12} \int dx dy \, \alpha(x) v(y) \, k_1  \{P_1(x),P_2(y)\}[k_2,\mu_2(y)] \nn \\
&& +  \text{tr}_{12} \int dx dy \, \alpha(x) v(y) \, k_1  [P_2(y),k_2] \{P_1(x),\mu_2(y)\}   \,.\nn
\eeqa
As in the previous calculation, we use the identity (\ref{intermediateresult}) and the defining relation of
the Casimir tensor to simplify this expression. This leads to:
\beqa
\{ H(\alpha) , H_0(v) \} &=& \text{tr}_{12}\int dx  \, \alpha(x) v(x) \, k_1
[t_{12},P_1(x)][k_2,\mu_2(x)] \nn \\
&&+  \text{tr}_{12}\int dx  \, \alpha(x) v(x) \, k_1[P_2(x),k_2][\mu_2(x),t_{12}]  \nn \\
&&-\text{tr}_{12}\int dx dy \, \alpha(x) v(y) \, k_1[P_2(y),k_2]t_{12} \, g_2(x)g_2^{-1}(y) \, \partial_y \delta(x,y) \nn \\
& = & \int dx  \, \alpha(x) v(x) \left( \text{tr}([P(x),k][k,\mu(x)] + \text{tr}([P(x),k][\mu(x),k] ) \right) \nn \\
 &&-\int dx  \, \alpha(x) v(y) \, \text{tr}(k[P(y),k]) \, \partial_y \delta(x,y) \,.\nn
\eeqa
The cyclicity of the trace, namely $\text{tr}(ab)=\text{tr}(ba)$, implies that the first two terms cancel and the last integral vanishes. As a result $\{ H(\alpha) , H_0(v) \}=0$.

\vspace{3mm}
$\bullet$ \underline{$\{ H_1(u) , H_1(u') \}$:}
\vspace{1mm}

\beqa
\{ H_1(u) , H_1(u') \} &=& \text{tr}_{12}\int dx dy \,u(x)u'(y) \, \{P_1(x)\mu_1(x),P_2(y)\mu_2(y)\} \nn \\
 & = & \text{tr}_{12}\int dx dy \,u(x)u'(y) \, \{P_1(x),P_2(y)\}\mu_1(x)\mu_2(y) \nn \\
 & & + \text{tr}_{12}\int dx dy \,u(x)u'(y) \, \{P_1(x),\mu_2(y)\}\mu_1(x)P_2(y)\nn \\
&& +\text{tr}_{12}\int dx dy \,u(x)u'(y) \,\{\mu_1(x),P_2(y)\}P_1(x)\mu_2(y).  \nn 
\eeqa
We replace the brackets by their expressions in the universal notations and after putting all the delta distributions
together and the derivatives of delta together as well, we obtain:
\beqa
\{ H_1(u) , H_1(u') \} & = & \text{tr}_{12}\int dx \,uu' \left([t_{12},P_1]\mu_1\mu_2 +\mu_1P_2[\mu_2,t_{12}]
-P_1\mu_2[\mu_1,t_{12}] \right)\nn \\
&& + \text{tr}_{12}\int dx dy \,u(x)u'(y) \, \left( P_1(x)\mu_2(y) \, \partial_x\delta(x,y) \right. \nn \\ && \left. -\mu_1(x) P_2(y)\, \partial_y\delta(x,y)\right)t_{12}.\nn
\eeqa
We have omitted to mention the $x$ or $y$ variables when it is not necessary. Then, it is easy to see that each of the three functions
appearing inside the first integral is given (up to a sign) by $\text{tr}([\mu,\mu],P)$ and therefore vanishes. As a consequence, the Poisson bracket reduces to the second integral which simplifies as follows:
\beqa
\{ H_1(u) , H_1(u') \} & = & \int dx dy \,u(x)u'(y) \, \left(\text{tr}(\mu(y) P(x)) \, \partial_x\delta(x,y) \right. \nn \\ && \left. - 
 \text{tr}(\mu(x)P(y)) \, \partial_y\delta(x,y)\right) \nn \\
& = & \int dx \, \left( -u'\text{tr}(\mu \partial_x(uP)) + u \text{tr}(\mu \partial_x(u'P))\right). \nn 
\eeqa
The second line is the result of an integration by part. The terms  in factor to the product $uu'$
cancel (which is in fact a result of the antisymmetry of the Poisson bracket) 
and the final result for the Poisson bracket we are interested in is:
\beqa
\{ H_1(u) , H_1(u') \} = \int dx \, (udu'-u'du) H_1(x) \, = \, H_1(udu'-u'du) \,.
\eeqa

\vspace{3mm}
$\bullet$ \underline{$\{ H_1(u) , H_0(v) \}$:}
\vspace{1mm}

\beqa
\{ H_1(u) , H_0(v) \} \, = \, \text{tr}_{12}\int dxdy \, u(x)v(y)\, \{P_1(x)\mu_1(x),P_2(y)[k_2,\mu_2(y)]\}. \nn
\eeqa
We proceed exactly as for the previous Poisson bracket. We first develop the bracket using the Leibniz rule,
then we use the defining relation of the Casimir tensor as well as the invariance of the trace, then we separate 
delta distributions from derivatives of delta and after some simple integrations, we get the right result.
All these steps are summarised in the following lines:
\beqa
\{ H_1(u) , H_0(v) \} & = & \text{tr}_{12}\int dxdy \, u(x)v(y)\, \{P_1(x)\mu_1(x),P_2(y)[k_2,\mu_2(y)]\} \nn\\
&=& \text{tr}_{12}\int dx \, u(x)v(x)\, \{t_{12},P_1(x)\}\mu_1(x)[k_2,\mu_2(x)] \nn\\
&& +\text{tr}_{12}\int dx \, u(x)v(x)\, \left( [\mu_2(x),t_{12}]\mu_1[P_2(x),k_2] \right. \nn \\ && \left. -P_1(x)[\mu_1(x),t_{12}][k_2,\mu_2(x)]\right)\nn\\
&& + \text{tr}_{12}\int dxdy \, u(x)v(y)\,P_1(x)t_{12}[k_2,\mu_2(y)] \, \partial_x\delta(x,y) \nn \\
&&- \text{tr}_{12}\int dxdy \, u(x)v(y)\, t_{12}\mu_1(x) [P_2(y),k]\,\partial_y\delta(x,y)\nn\\
&=& \int dx \, u(x)v(x)\, \text{tr}([[k,\mu],P]\mu +[\mu,\mu][P,k]-P[\mu,[k,\mu]])\nn \\
&&+\int dxdy \, u(x)v(y)\,\text{tr}(P(x)[k,\mu(y)])\,\partial_x\delta(x,y)\nn\\
&&-\int dxdy \, u(x)v(y)\,\text{tr}(\mu(x)[P(y),k])\,\partial_y\delta(x,y).\nn
\eeqa
The first integral (in the last equality) vanishes and only remain the last two integrals
involving derivatives of delta:
\beqa
\{ H_1(u) , H_0(v) \} \, = \, H_0(udv-vdu)\,.
\eeqa

\vspace{3mm}
$\bullet$ \underline{$\{ H_0(v) , H_0(v') \}$:}
\vspace{1mm}

\beqa
\{ H_0(v) , H_0(v') \} &=& \text{tr}_{12}\int dxdy \, v(x)v'(y)\, \{P_1(x)[k_1,\mu_1(x)],P_2(y)[k_2,\mu_2(y)]\}. \nn
\eeqa
Using the Leibniz rule and the cyclicity of the trace, the Poisson brackets reads: 
\beqa
\{ H_0(v) , H_0(v') \} & = & \text{tr}_{12}\int dxdy \, v(x)v'(y)\, \{P_1(x),P_2(y)\} \, [k_1,\mu_1(x)]\,[k_2,\mu_2(y)] \nn \\
&&+\text{tr}_{12}\int dxdy \, v(x)v'(y)\, \{P_1(x),\mu_2(y)\}\, [k_1,\mu_1(x)]\,[P_2(y),k_2] \nn \\
&&+\text{tr}_{12}\int dxdy \, v(x)v'(y)\, [P_1(x),k_1] \, \{\mu_1(x),P_2(y)\} \,[k_2,\mu_2(y)].\nn
\eeqa
We replace each Poisson bracket by their expressions which involve delta distributions and derivatives of delta
 distributions. We separate the deltas from their derivatives and we obtain:
\beqa
\{ H_0(v) , H_0(v') \} & = & \text{tr}_{12}\int dx \, v(x)v'(x) \, [t_{12},P_1][k_1,\mu_1][k_2,\mu_2] \nn \\
&&+ \text{tr}_{12}\int dx \, v(x)v'(x) \,\left([\mu_2,t_{12}][k_1,\mu_1][P_2,k_2] \right. \nn \\ && \left.
-[P_1,k_1][\mu_{1},t_{12}][k_2,\mu_2] \right)\nn\\
&&+\text{tr}_{12}\int dx dy\,v(x)v'(y) \, [P_1(x),k_1]t_{12}[k_2,\mu_2(y)]\,\partial_x\delta(x,y)\nn \\
&&-\text{tr}_{12}\int dx dy\, v(x)v'(y) \,t_{12}[k_1,\mu_1(x)][P_2(y),k_2]\,\partial_y\delta(x,y). \nn
\eeqa
Using the defining property of the Casimir tensor, we show that the first line can be written as an integral of 
the product of $vv'$ with the function $\text{tr}([[k,\mu],P][k,\mu])$ which vanishes due to the invariance of the trace.
The second line is a sum of two terms, each of them being an integral of the product of $vv'$ with 
the function $\text{tr}([\mu,[k,\mu]][P,k])$ or $-\text{tr}([\mu,[k,\mu]][P,k])$. Therefore, the second line is identically
null as well. Only the terms involving the derivatives of delta remain as expected from the antisymmetry of the Poisson
bracket:
\beqa
\{ H_0(v) , H_0(v') \} & = & \int dx dy\,v(x)v'(y) \, \text{tr}([P(x),k][k,\mu(y)])\,\partial_x\delta(x,y)\nn\\
&& - \int dx dy\,v(x)v'(y) \, \text{tr}([P(y),k][k,\mu(x)])\,\partial_y\delta(x,y).\nn
\eeqa
We perform an integration by part with respect to the $x$ and $y$ variables respectively
in the first and second integral and after the integration of the delta distribution, we obtain:
\beqa
\{ H_0(v) , H_0(v') \} \, = \,  \int dx \,(vdv'-v'dv)\text{tr}([P(x),k][k,\mu(x)]).\nn 
\eeqa
Next, we use the fact that $[a,[b,c]]=\text{tr}(ac)b-\text{tr}(ab)c$ for any $\so(3)$ elements $a$, $b$ and $c$.
In the case where $a=k$, $b=k$ and $c=[k,\mu]$,  this identity becomes $[k,[k,\mu]]=\text{tr}(k\mu)k-\text{tr}(k^2)\mu$
and leads to the equality:
$$
\text{tr}([P(x),k][k,\mu(x)]) \, = \, \text{tr}(P(x)k)\text{tr}(k\mu(x)) \, - \, \text{tr}(k^2)\text{tr}(P(x)\mu(x)).
$$
This identity allows to simplify the Poisson bracket between the constraints $H_0$ according to:
\beqa
\{ H_0(v) , H_0(v') \} \, = \, -\text{tr}(k^2)\,H_1(vdv'-v'dv) \, + \, \tilde{H}(vdv'-v'dv),
\eeqa
where $\tilde{H}(x)=\text{tr}(P(x)k)\text{tr}(k\mu(x))=H(x)\text{tr}(k\mu(x))$ by definition.
Finally, the relation $(2\pi\alpha')^2\text{tr}(k^2)=1$ between $k$ and the Regge slope $\alpha'$ leads
to the expression of the Poisson bracket announced in the theorem.

This closes the proof of the theorem \ref{symmetry} $\square$.

\medskip

We close this section with some remarks.
Firstly, although algebraically closed, the symmetry algebra of the self-dual algebraic string computed above is open in the BRST sense, that is, we have structure functions instead of structure constants because the smearing field in the bracket between $H_0$ and $H_0$ depends on the dynamical fields themselves. This prevents the constraints from forming an infinite dimensional Lie algebra, exactly like in the Ashtekar (and Ashtekar-Barbero-Immirzi) formulation of general relativity (see for instance \cite{giulini} and references therein). 

\medskip

This leads us to our second remark. It is important to emphasise the analogies between the symmetry algebra of the self-dual algebraic string and the Poisson algebra of the first class constraints appearing in Ashtekar gravity.
Indeed, if we interpret $H_1$ and $H_0$ as the analogues of the gravitational vector (or diffeomorphism) and scalar (or Hamiltonian) constraints respectively, and $H$ as the analogue of the Gauss constraint - which is consistent because both constraints do not appear in the second order, metric formulation and reflect the freedom in choosing the first order variable (the frame field or $g$) - both algebras are identical. Indeed, in gravity the Gauss constraint generates an ideal of the symmetry algebra; the Gauss constraint closes with itself, yields Gauss when commuted with the vector constraint and vanishes with the scalar constraint, while the diffeomorphism constraint only generates a subalgebra because it closes with itself but yields the Hamiltonian constraint when commuted with the latter constraint. Finally, the bracket between two scalar constraints produces a vector constraint augmented with a field dependent Gauss constraint. All these features appear in the symmetry algebra of the self-dual algebraic string and the two algebras, modulo the proposed identifications, are therefore exactly identical.

\medskip

Finally, it is important to compare the symmetry algebra of the self-dual algebraic string with the Witt algebra of diffeomorphisms on the world sheet of ordinary string theory. The commutation relations involving $H_1$ and $H_0$ fail to reproduce exactly the Witt algebra because of the extension $\tilde{H}$ appearing in the Poisson bracket between $H_0$ and itself. Furthermore, the Witt algebra is not even a sub-algebra of the symmetry algebra. This could seem puzzling at first. In fact, the answer to this paradox is rather simple: when restricted to the functions $f$ invariant under the action of $H$, i.e. $\{H,f\}=0$, the symmetry algebra reduces exactly to the Witt algebra as expected. In other words, we recover the Poisson algebra  of constraints of ordinary string theory when the symmetry generated by $H$ has been factored out. Note that this separate phase space reduction is always possible because the phase space flow generated by $H_1$ and $H_0$ leaves the $H=0$ hypersurface invariant.

\subsubsection{Symmetries and Dynamics}

This section is devoted to compute the action of the constraints on the configuration variables of the theory.
These calculations will not only make clear the interpretation of the symmetries but will also allow to discuss
and eventually construct the classical physical phase space. In a second part, we will exhibit the dynamics
and show that the associated equations of motion for the configuration embedding variable are the wave equations as expected.

The action of the symmetries on the dynamical variables are given in the following lemma.

\begin{lem}
\label{lemmesym}
The action of the smeared constraints $H(\alpha)$, $H_1(u)$, and $H_0(v)$ on the configuration variables $(X,g)$ is displayed by the following expressions. The action of $H$ yields
\beq
\delta_{\alpha} X^I = \{ H(\alpha) , X^I \}_D = 0, \;\;\;\; \;\;\;\; \delta_{\alpha} g = \{ H(\alpha) , g \}_D = -\alpha kg,
\eeq
while $H_1$ induces the transformations
\beq
\delta_{u} X^I = \{ H_1(u), X^I \}_D = - u d X^I, \;\;\;\; \;\;\;\; \delta_{u} g = \{ H_1(u) , g \}_D = - u d g.
\eeq
Finally, $H_0$ generates the following action
\beq
\delta_{v} X^I = \{ H_0(v) , X^I \}_D = 2 v M^I{}_J d X^J, \;\;\;\; \;\;\;\; \delta_{v} g = \{ H_0(v) , g \}_D = - v g d M \,.
\eeq
\end{lem}

{\em Proof.} Firstly, we consider the general action of the constraints $(\phi_a)_a$ on the configuration variables. The Dirac brackets with the embedding variables yields
\beq
\{ \phi_a(x) , X^I(y) \}_D = - \frac{1}{2} (D(x)^{-1})^{IJ} \left(g(x)^{-1}[T_{a} , k] g(x) \right)_{JK} d X^K(x) \, \delta(x,y),
\eeq
while the action on the group elements produces
\beq
\{ \phi_a(x) , (g(y))^I_{\;J} \}_D = - (T_{a} \, g(x))^I_{\;J} \, \delta(x,y).
\eeq
From these preliminary computations, it is immediate to calculate the action of the three smeared first class constraints $H = \phi(k)$, $H_1 = \phi(\mu)$ and $H_0 = \phi([k,\mu])$ on the dynamical variables. The first constraint yields
$$
\{ H(\alpha) , X^I \}_D = 0, \;\;\;\; \;\;\;\; \{ H(\alpha) , g^I_{\;J} \}_D = - \alpha\, (kg)^I_{\;J}.
$$
The second Poisson bracket has been computed using the matrix elements of $g$ and therefore extends in the universal
notation to the group element itself. We proceed similarly for the other cases.
The second calculation leads to
$$
\{ H_1(u) , X^I \}_D = - u d X^I, \;\;\;\; \;\;\;\; \{ H_1(u) , g^I_{\;J} \}_D =  - u dg^I_{\;J},
$$
where we have used equation \eqref{com} to recognise the exact inverse of the Dirac matrix to obtain the first equality. Finally, using the relation between the Dirac matrix and the derivative of $M$, equation \eqref{com}, and the fact that $dM M = - M d M$ (because $M^2$ is a constant matrix), one obtains that the action of the last constraint 
$$
\{ H_0(v) , X^I \}_D = 2 v M_{\; J}^I d X^J, \;\;\;\; \;\;\;\; \{ H_0(v) , g^I_{\;J} \}_D = - v (g d M)^I_{\;J}.
$$
Thus, lemma $2$ is proved $\square$.

\medskip

From the above lemma, we can immediately conclude that $H(\alpha)$ generates infinitesimal left $\SO(2)$ transformations stabilising $k$ with parameter $\alpha$, and that $H_1(u)$ generates infinitesimal diffeomorphisms of the circle with vector field $u \, \partial_x$. By elimination, the last first class constraint $H_0(v)$ out of the three necessarily generates infinitesimal diffeomorphisms in the time-like direction, i.e., time reparametrisation. 

To confirm this last point, we can compute the equations of motion for $X$ associated to the Hamiltonian $H_0$. The evolution of the dynamical variables are encoded in the following brackets
\beqa
\partial_t X^I &:=& \{H_0 , X^I \}_D = 2 v M_{\; J}^I \partial_x X^J, \\
\partial_t g^I_{\;J} &:= &\{H_0 , g^I_{\;J} \}_D = - v (g \partial_x M)^I_{\;J}.
\eeqa
From the second equation, we can calculate $\partial_t M = [\partial_x M, M]$, and consequently compute the second time derivative of the $X$ variable 
\beqa
\partial_t^2 X^I &=& 2 v \partial_t M_{\; J}^I \partial_x X^J + 2 v M_{\; J}^I \partial_t \partial_x X^J \nn \\
                 &=&  2 v^2 [\partial_x M,M]^I_{\; J} \partial_x X^J +4 v M_{ \; J}^I \partial_x (v
M_{\; K}^J
\partial_x X^K) \nn \\
                 &=& 2 v^2 \left((\partial_x M \, M)^I_{\; J} + (M \, \partial_x M)^I_{\; J} \right) 
\partial_x X^J \nn \\ 
&& + 4 v^2 (M^2)^I_{\; J} \, \partial_x^2 X^J + 4 v \partial_x v (M^2)^I_{\; J} \, \partial_x X^J.
\eeqa
Using again the fact that $M^2 = - 1 / (2\pi \alpha')^2 1\!\! 1$ is a constant matrix, we obtain the standard Euclidean-covariant wave equation for $X^I$
\beq
\left(\partial_t^2 + \partial_x^2 \right) \, X^I = 0,
\eeq
upon the choice of gauge $v = \pi \alpha'$. This result ensures that $H_0$ can be interpreted as the Hamiltonian
constraint of the theory.

Let us emphasise that we have a freedom to choose the dynamics of the theory due to the
diffeomorphisms invariance. One could have defined the time derivative from a ``Hamiltonian'' $H_t$ of the form
$H_t=H_1(u)+H_0(v)$ for any smooth functions $u$ and $v$ as soon as $v \neq 0$. It is quite interesting to note that
the very natural choice $u=0$ and $v=cte$ leads to the wave equation as the equation of motion of 
the string coordinates. Interestingly, in the context of the Polyakov string, this equation is recovered with exactly the same choice of gauge fixing.

\subsubsection{Physical degrees of freedom}

So far, we have an implicit description of the physical phase space only. As we have already mentioned, a direct 
calculation
shows that the theory admits 4 degrees of freedom (in the phase space) as expected: 
indeed $2\times4$ variables $(X^I,\pi_I)$ with 
$2\times 3$ variables $(g,P)$ supplemented with 4 second class constraints and 3 first class constraints lead to
$2\times 4+ 2\times 3 - 4 -2\times 3=4$ degrees of freedom.

As we have explicitly computed the Dirac bracket, we can set the second class constraints to zero. The consequence is that
we can forget the variables $\pi_I$ which are explicitly given in terms of $X^I$ and $g$ and 
we can also set the first class constraints $P$ to zero. Finally, we are left with the variables $X^I$ and $g$
satisfying the Poisson algebra
$$
\{g_1,g_2\}_D\,=\,0 \;,\;\;\;\;
\{g,X^I\}_D\,=\,0 \;,\;\;\;\; \{X^I,X^J\}_D\,=\,(D^{-1})^{IJ}
$$
with the symmetry actions described in (\ref{lemmesym}) as automorphisms of this algebra. This description allows to
simplify considerably the definition of the physical phase space $\cal P$ which is now symbolically constructed as follows:
$$
{\cal P} \, := \, {\cal F}(X,g)/{\text{Sym}}
$$
where $\cal F$ denotes the set of smooth functions on the configuration variables and $\text{Sym}$ is for 
the symmetry action induced on $\cal F$. The space is endowed with a non-degenerate symplectic structure and
admits of course $4 (X^I \, \text{variables}) +3 (g \, \text{variables}) -3(\text{constraints})\,=\,4$ degrees of freedom.

A complete description of the phase space would come with a parametrisation of it. We will not give here a precise
parametrisation of the diffeomorphisms orbits (generated by $H_1$ and $H_0$) for this question has been studied deeply
in the context of the bosonic string for instance. We will rather parametrise the orbits generated by $H$ (hereafter called
$H$-orbits)
which is much
simpler to do. Indeed, $H$ does not affect the $X^I$ variables and acts as a right derivative on the group variable $g$.
It is immediate to see that the set of $H$-orbits is simply given by the conjugacy class of the element $k$
which is parametrised by the elements $M=g^{-1}kg$. As a consequence, if we work in phase space with the variables $M$ instead of $g$
we can forget the constraint $H$ and set it explicitly to zero. Then, the physical phase space is symbolically
constructed as the coset:
$$
{\cal P} \, := \, {\cal F}(X,M)/{\text{Diff}}
$$
where the symmetry algebra is reduced to the algebra of diffeomorphisms on the world sheet, denoted $\text{Diff}$
above.
This definition makes sense because $\text{Diff}$ leaves the set of $H$-orbits invariant.
Finally, the algebraic string provides a new description of the physical phase space of the bosonic string. 
The major novelty with this description is that the non-reduced phase space, namely ${\cal F}(X,M)$, is 
such that the string coordinates are non-commutative variables, contrary to what happens with the Polyakov
or the Nambu-Goto string. As for the group variables $g$, they can be interpreted as central extension of the 
non-commutative algebra. This interpretation is nonetheless misleading because they transform non-trivially
under the gauge symmetries. 

To finish the classical study of the self-dual algebraic string, let us try to explain the differences
and to clarify the link between the physical phase space $\cal P$ and the one inherited from the Nambu-Goto or 
Polyakov action. First of all, the two symplectic spaces are isomorphic and then they are the same. In the 
Polyakov-Nambu-Goto string, the physical phase space would be explicitly constructed if not only one solves the
diffeomorphisms constraints but also one finds the gauge orbits. In the algebraic string, the physical phase space
is obtained once one finds the gauge orbits only. In a sense, the constraints have already been solved in that context.
The non-commutativity of the string coordinates is the price to pay, so to say, to have solved the constraints. 

\subsection{General case: arbitrary Immirzi-like parameter}

As  explained in section 2, adding a non-trivial Immirzi-like parameter $\gamma$ in the algebraic string context reduces to a particular choice of $k\in \so(4)$. The string tension and the Immirzi parameter are related
to $\langle k,k \rangle$ and $(k,k)$ in a simple way (Section 2.2.4.). This general formulation covers the original framework as proposed in \cite{Bal1} by setting the Immirzi-like parameter to zero. Hence, a Hamiltonian analysis of the algebraic string in this general context will describe the canonical aspects of the original proposal ($\gamma=0$) and also the classical effect of an Immirzi like parameter. This is the purpose of this section.

We will proceed as in the self-dual case: we will extract the constraints, separate the first class from the second class constraints, compute the Dirac bracket and so on so forth. The calculations are very similar to the previous ones; for that reason, we will not focus on the technical aspects. We will rather point the main differences with the self-dual case. In particular, we will show that, contrary to what happens in gravity, the classical non-self-dual theory ($\gamma \neq 1$) is drastically different from the self-dual formulation. Indeed, a new degree of freedom appears in the configuration space. This prevents the generic algebraic string from being equivalent to the Nambu-Goto string. However, there are very interesting similarities with gravity. For instance, the introduction of an Immirzi-like parameter modifies only the expression of the Hamiltonian constraint $H_0$ keeping the vectorial constraint $H_1$ and the "Gauss like" constraint $H$ unchanged compared to the self-dual case.

\subsubsection{Number of physical degrees of freedom}

The starting point is the action (\ref{general}) where $M=g^{-1}kg$ is not restricted to any sub-algebra of $\so(4)$.
In particular, $\langle k,k \rangle  \pm (k,k) \neq 0$ which prevents to restrict ourselves to the self-dual or 
anti-self-dual cases. 

We start with the same Poisson brackets as in the self-dual case (Section 3.1.1.) with the difference that 
the group element $g$ is now an element of $\SO(4)$ and therefore the associated momentum $P_A$ is a 6 dimensional
vector. The index $A$ labels a basis of the algebra $\so(4)$: either it can be viewed as a pair of integer $(ab)$
with $a < b =0,...,3$, either as a couple $(\epsilon,a)$ with  $\epsilon =\pm$ and  $a=1,2,3$.

The primary constraints are the same 4 constraints $C_I=\pi_I-\frac{1}{2}M_{IJ}dX^J\approx 0$ 
supplemented with the 6 constraints $\phi_A=P_A\approx 0$. There are no secondary constraints
and therefore there are 10 constraints in total that we have to decompose into first class and
second class. Before going into the details of the constraints analysis, let us give a brief
summary of our results: among the 10 constraints, 4 are first class and then the 6 remaining are second class.
Among the first class constraints, we recover the diffeomorphisms and the interpretation of the two others
will be given in the sequel.
As we have started with $2\times 4+ 6\times 2=20$ degrees of freedom in the phase space,
we conclude that the theory possesses $20\, (degrees \,of \,freedom)-6\, (second \, class, constraints)-2\times 4
\, (first \, class, constraints)=6$ physical degrees of freedom in the phase space. Two more than the Nambu-Goto
or the Polyakov string. As a consequence, if not restricted to the self-dual or anti-self dual cases, the
algebraic string is not equivalent, even classically, to the standard bosonic string. 
However, from a naive Lagrangian analysis, we see that the algebraic string contains the bosonic string as a solution.
This remark raises 
many questions that we hope to answer in the future: what is the status of the extra degree of freedom?
In which way  this extra degree of freedom is coupled to the Nambu-Goto string?

\subsubsection{Partial Dirac bracket}

The canonical way to distinguish the first class from the second class constraints consists in first
computing the constraints matrix and then finding its kernel. The kernel is generated by the first class
constraints. This method is systematic but often quite fastidious. Here, we will proceed in a recursive way:
first, we consider a subset of second class constraints; then, we compute the associated Dirac bracket
which implies that we can eliminate explicitly these second class constraints; finally, we are left with a
smaller system of constraints and repeat the method until there is no more second class constraint. At the end
of the process, we have separated the first class from the second class constraints.

We start with the subset of constraints generated by $C_I$. As in the self-dual case,
we can show that this subset is second class. Indeed, a similar calculation leads to (symbolically) the same Dirac 
matrix (associated to the constraints $C_I$) as in the self-dual case, namely $D_{IJ}=-\frac{1}{2}dM_{IJ}$.
The major difference is that $M$ is an $\so(4)$ matrix not restricted to the self-dual sub-algebra $\so(3)_+$.
However, the Dirac matrix can be decomposed into self-dual and anti-self-dual components according to \eqref{iso}:
$$
D_{IJ} \; = \; D_{+IJ} \, - \, D_{-IJ}.
$$
The two components commute, $D_+D_-=D_-D_+$, and satisfy the relation (\ref{squarerelation}):
$$
D^2_\pm \; = \; -\frac{1}{4}\text{tr}(D^2_\pm)\, 1\!\!1\,.
$$
As a consequence, $D$ is invertible if $\text{tr}(D^{2}_\pm)\neq 0$ (which is trivially the case as soon as
$M$ is not restricted to lie in the self-dual or anti-self-dual sectors) and its inverse is simply obtained
from the relation
\beq
\label{D}
D \star D \, = \, (D_+ - D_-)\, (D_+ + D_-)\, = \, D^{2}_+ - D^{2}_- \, = \,  
-\frac{1}{2} \la D, \star D \ra \, 1\!\!1\, ,
\eeq
as follows
\beqa
D^{-1} \, = \, -\frac{2}{(D^2)} (D_+ + D_-) \, = \,
-\frac{2}{(D^2)} \star D \,.
\eeqa
Therefore, the set of constraints $\{C_I, I=1,\cdots,4\}$ is of second class as expected. One can compute the
associated Dirac bracket from the expression of $D^{-1}$ and eliminate explicitly the constraints $C_I$. The obtained
bracket is called a partial Dirac bracket because, so far, we do not know if second class constraints are remaining
in the system. This is what we want to analyse now.

\subsubsection{First class constraints}

The (partial) Dirac brackets between the remaining six constraints $\phi_A$ are very similar to the ones computed in
the self-dual case (\ref{dirac}):
$$
\{\phi_A(x),\phi_B(y)\}_D \, = \, \{\phi_A(x),\phi_B(y)\} \, - \, \Delta_{ABIJ}(x)dX^IdX^J \,\delta(x,y)
$$
where the extension $\Delta_{ABIJ}(x)$ reads (symbolically):
\beqa\label{expressdelta}
\Delta_{ABIJ}(x) & = & \frac{1}{4}\{\phi_A(x),M_{KI}\}(D^{-1})^{KL}\{M_{LJ},\phi_B(x)\} \nn\\
&=& -\frac{1}{2 (D^2)}\{\phi_A(x),M_{KI}\}\star D^{KL}\{M_{LJ},\phi_B(x)\} \nn \\
&=& -\frac{1}{4 (D^2)} \left(g^{-1}[T_A,k][dgg^{-1},\star k][T_B,k]g \right)_{IJ}\,.
\eeqa
We used the relation $D=\frac{1}{2}g^{-1}[dgg^{-1},k]g$ and the relation $\star[a,b]=[\star a,b]$
for any Lie algebra elements $a$ and $b$ in $\so(4)$.

The argument we gave to show that the `central extension' $\Delta$ vanishes in the self-dual case does not work anymore and we cannot conclude that all the six constraints $\phi_A$ are first class. To exhibit the first class
constraints out of the six, it is convenient to introduce a local basis of the set of constraints.
This new basis is very similar to the one defined
in the self-dual case (\ref{localbasis}) and consists into the following:
\beqa
&&\Gamma:=\phi(k) ,\;\;\;\; \Gamma_1:=\phi(\mu),\;\;\;\;\Gamma_0 := \phi([k,\mu]) \nn \\
&&\Gamma^\star:=\phi(\star k) ,\;\;\;\; \Gamma^\star_1:=\phi(\star \mu),\;\;\;\;\Gamma^\star_0 := 
\phi([\star k,\mu])\,.
\eeqa
In the self-dual case, we would have $\Gamma=\Gamma^\star$ and $\Gamma_i=\Gamma_i^\star$ for $i=0,1$ but the constraints are  independent in the general case. Furthermore, they allow to find quite easily the first class constraints even if all 
of them are not given explicitly.

\begin{theo}[First class constraints.]
Among the six remaining constraints, four are first class. Three of them are given by $\Gamma$, $\Gamma^\star$
and $\Gamma_1$. The fourth one is given by the one-dimensional kernel of the three dimensional
constraints matrix constructed from $\Gamma_0$, $\Gamma_1^\star$ and $\Gamma_0^\star$. It is (weakly) equal to
\beqa\label{hamiltconstraint}
\left((\star \mu)^A (\star[k,\mu])^B \Gamma_0 - [k,\mu]^A(\star[k,\mu])^B \Gamma^\star_1+ 
[k,\mu]^A(\star \mu)^B \Gamma^\star_0\right) \Delta_{ABIJ}dX^IdX^J
\eeqa
where $a^A$ is the component of the Lie algebra element $a=a^AT_A$ in the basis $(T_A)_A$.
\end{theo}

{\em Proof.}
A linear combination of the constraints $\Phi:=v^A\phi_A$ is first class with respect to the Dirac bracket
if its Dirac bracket with all the constraints $\phi_A$ vanish even weakly. This condition  is
satisfied if
$$
v^A\Delta_{AB(IJ)}(x) \; = \; 0, 
$$
where $(IJ)$ denotes the normalised symmetrisation of the tensor. Let us show that this relation is true for the three constraints
$\Gamma$, $\Gamma^\star$ and $\Gamma_1$.

\vspace{3mm}
$\bullet$ \underline{The constraint $\Gamma$:}
\vspace{1mm}

The vector $v$ associated to $\Gamma$ is $v=k$. Hence, it is immediate to see that
$$
k^A\Delta_{ABIJ}(x) \; = \; -\frac{1}{4 (D^2)} \left(g^{-1}[k,k][dgg^{-1},\star k][T_B,k]g \right)_{IJ}\; = \; 0
$$
due to the presence of the commutator $[k,k]=0$. This proves that $\Gamma$ is first class.

\vspace{3mm}
$\bullet$ \underline{The constraint $\Gamma^\star$:}
\vspace{1mm}

A very similar argument works to show that $\Gamma^\star$ is a first class constraint as well.
In that case, we find that
$$
\star k^A\Delta_{ABIJ}(x) \; = \; -\frac{1}{4 (D^2)} \left(g^{-1}[\star k,k][dgg^{-1},\star k][T_B,k]g \right)_{IJ}\; = \; 0
$$ 
because $k$ and $\star k$ commute.

\vspace{3mm}
$\bullet$ \underline{The constraint $\Gamma_1$:}
\vspace{1mm}

Proving that $\Gamma_1$ is first class is a bit more subtle. The vector $v$ associated to this constraint is 
$dg g^{-1}$ and we have
$$
(dg g^{-1})^A\Delta_{ABIJ}(x) \; = \; -\frac{1}{4 (D^2)} \left(g^{-1}[dg g^{-1},k][dgg^{-1},\star k][T_B,k]g \right)_{IJ}\,.
$$
To show that this quantity vanishes, we first observe that the product $a\star a$ is proportional to the identity for any $\so(4)$ Lie algebra element $a$. The proof has been given in for the Dirac matrix in \eqref{D} but applies to all $\so(4)$ elements and shows that $a \star a= -\frac{1}{2} (a^2) 1 \!\! 1$.
We see that the product $[dg g^{-1},k][dgg^{-1},\star k]=[dg g^{-1},k]\star[dgg^{-1},k]$ appears in the expression of
$(dg g^{-1})^A\Delta_{ABIJ}(x)$ which then simplifies as follows
$$
(dg g^{-1})^A\Delta_{ABIJ}(x) \; = \; \frac{1}{8 (D)^2} \, ([dg g^{-1},k]^2) \,
\left(g^{-1}[T_B,k]g \right)_{IJ}\,.
$$
As a consequence, the symmetrised tensor $(dg g^{-1})^A\Delta_{AB(IJ)}(x)=0$ because $a_{(IJ)}=0$ for any Lie algebra 
element $a$. This proves that $\Gamma_1$ is a first class constraint.

\vspace{3mm}
$\bullet$ \underline{The remaining first class constraint:}
\vspace{1mm}

At this stage, we are left with three remaining constraints $\Gamma_0$, $\Gamma_1^\star$ and $\Gamma_0^\star$.
The associated constraints matrix $C$ with respect to the Dirac bracket is then three dimensional. 
It is defined as follows:
\beqa
C:=\left(
\begin{array}{ccc}
0 & \{\Gamma_0,\Gamma^\star_1 \}_D & \{\Gamma_0,\Gamma^\star_0 \}_D \\
\{\Gamma^\star_1,\Gamma_0 \}_D & 0 & \{\Gamma^\star_1,\Gamma^\star_0 \}_D \\
\{\Gamma^\star_0,\Gamma_0 \}_D & \{\Gamma^\star_0,\Gamma^\star_1 \}_D & 0
\end{array}
\right) 
\eeqa
The expression (\ref{expressdelta}) of the Dirac bracket between the constraints leads to:
\beqa
C \approx -\Delta_{ABIJ}dX^IdX^J\left(
\begin{array}{ccc}
0 & [k,\mu]^A (\star \mu)^B& [k,\mu]^A (\star[k,\mu])^B  \\
(\star\mu)^A [k,\mu]^B & 0 &  (\star \mu)^A (\star[k,\mu])^B\\
 (\star[k,\mu])^A [k,\mu]^B & (\star[k,\mu])^A  (\star \mu)^B & 0
\end{array}
\right) \nn
\eeqa
We recall that $\approx$ denotes the weak equality.
As the number of
second class constraints is always odd, either two out of the three constraints are second class or no one of them.
However, it is easy to check that the matrix constraint is not weakly vanishing. Therefore, two out of the three
constraints are second class. The kernel of the constraints matrix gives the first class constraint 
(\ref{hamiltconstraint}) $\square$.

\medskip

This theorem leads to important remarks. Firstly, the question of the physical interpretation of the first class constraints needs to be addressed. To clarify this point,
it is necessary to compute the action of the constraints on the configuration variables with respect
to the Dirac bracket. Because the calculations are the same than in the self-dual case, we will not give the 
technical details. The results are the following. 

The constraints $\Gamma$ and $\Gamma^\star$ have a trivial action on the world sheet coordinates $X^I$ whereas they act on the group variable $g$ by a left multiplication respectively by the element $k$ and $\star k$. We recover here the symmetries we have easily observed in the Lagrangian framework. 
Indeed, the group variable appears only via $M=g^{-1}kg$ in the Lagrangian and $M$ is clearly invariant
under the action of $\Gamma$ and $\Gamma^\star$.

The constraint $\Gamma_1$ has formally the same structure than the vectorial constraint $H_1$ in the self-dual case.
It is then natural to expect that it is the generator of space diffeomorphisms in the general context.
To verify this is indeed the case, let us compute its action
on the world sheet coordinates $X^I$:
\beqa
\{\Gamma_1,X^I\}_D \, &=& \, -\{\Gamma_1,C_K\}(D^{-1})^{ KI} \nn \\
&=& - \frac{1}{2} (g^{-1}[\mu,k]g)_{KL} (D^{-1})^{ KI} dX^L.  \nn
\eeqa
Recognising the matrix $D$ in disguise, we immediately conclude that
$$
\delta_1 X^I \, = \, \{\Gamma_1,X^I\}_D \, = \, dX^I\,
$$
which confirms the interpretation of $\Gamma_1$ as the vectorial constraint. Finally, it is natural
to consider the last first class constraint as the scalar constraint that generates the dynamics.

\medskip

As a second remark, let us again emphasise the similarities with the gravitational case. Only the expression of the scalar constraint formally differs from the self-dual case. The vectorial constraint $\Gamma_1$ has
the same expression as $H_1$. The same observation is true for the Gauss like constraints $\Gamma$ and $\Gamma^*$ which are similar to $H$.

\medskip

The last remark concerns the expression of the Hamiltonian constraint.
It is immediate to give an explicit form for the Hamiltonian constraint but we do not have
a simple formula for it. Therefore, it is not immediate to see where the Immirzi-like parameter
appears in the Hamiltonian constraint. This point is crucial to clearly understand the classical and quantum effects of this parameter and is currently under investigation.

\section{Conclusion}
In this work, we have studied the Hamiltonian analysis of the algebraic string.
The algebraic string was introduced more than two decades ago by Balachandran and
collaborators as a first order formulation of the Nambu-Goto string. At the Lagrangian level,
these two string formulations seem to be equivalent. However, the situation is more subtle.
A careful Hamiltonian analysis shows that this is not generically the case. The equivalence is true only in the self-dual and anti-self-dual sectors, as shown in this article.

Indeed, we have discovered that the algebraic string admits, as in general relativity, a self-dual
formulation and an Immirzi-type parameter. We have done the canonical analysis of the system in the two cases.
The self-dual string has been shown to be equivalent to the standard Nambu-Goto string but has lead us to a new formulation of the physical phase space. Indeed, in this framework, the world sheet coordinates are non-commutative once we solve the second class constraints computing the Dirac bracket. As expected, the first class constraints generate the diffeomorphisms on the world sheet and act as automorphisms on the phase space.

In the non-self-dual case, covering both the original proposal and the inclusion of a non-trivial Immirzi-type parameter, we showed that the system admits one more
degree of freedom in the configuration space than the two of the Nambu-Goto string. This prevents the theory from being strictly equivalent to the standard bosonic string. Nonetheless, we exhibited the first class constraints
which generate the diffeomorphisms on the world sheet, namely the scalar and the vectorial constraints.

\medskip

One of the most strinking aspects of the algebraic string is its numerous similarities
with the frame formulation of gravity. It is a first order theory, it is of course diffeomorphism invariant, and admits a self-dual formulation and an Immirzi type parameter. All these aspects makes the system a very nice arena to test the ideas and techniques of LQG because it is simple enough
to be completely quantised. The Fock quantisation already partially exists. Our aim, in future work, 
is to develop a background independent quantisation \`a la LQG in order to conclude on the equivalence or not with
the Fock quantisation. This idea was in fact initiated by Thiemann \cite{Thiemann} in the context of the Nambu-Goto 
string but we think that the algebraic formulation of the string is more suited for that specific problem.
Furthermore, there is an Immirzi-type parameter and then we hope to understand its effects in the quantum theory. 
We hope the algebraic string helps us to understand some other fundamental aspects of LQG.
For instance, we can address the question of the existence of 
a spinfoam \cite{SF} formulation of the algebraic string. If this was the case, we would have a new arena, simpler than gravity, to understand the link between the covariant and canonical quantisations of background independent theories.


\subsubsection*{Acknowledgements}
We would like to thank Xavier Bekaert for his interest and some clarifying discussions about
the bosonic string. K.N. wants to thank Jihad Mourad and Giuseppe Policastro for discussions
and for their interest on the subject.
W.F. is supported by the Royal Commission for the Exhibition of 1851 and would like to thank the Laboratoire de Math\'ematiques et de Physique Th\'eorique (LMPT)
of Tours for their hospitality. This work was partially supported by the ANR ($BLAN06-3\_139436 LQG-2006$).

\end{document}